\documentclass[reprint, amsmath,amssymb, aps, prb, floatfix, ]{revtex4-2}

\usepackage{graphicx} 
\usepackage{multirow}
\usepackage{amsfonts}
\usepackage{dsfont}
\usepackage{braket}
\usepackage{xcolor}
\usepackage{ulem}
\usepackage{comment}
\usepackage{natbib}
\usepackage{hyperref}
\hypersetup{colorlinks=true, linkcolor=blue, filecolor=blue,   citecolor=blue,   urlcolor=blue     }

\begin{document}

\preprint{APS/123-QED}

\title{Interplay of frustration and quantum fluctuations in a spin-1/2 anisotropic square lattice}

 \author{L. M. Ramos}%
 \email{lucas.morais@ufms.br}
 
\author{F. M. Zimmer}%
 \email{fabio.zimmer@ufms.br}
\affiliation{%
 Instituto de Física, Universidade Federal de Mato Grosso do Sul, 79070-900, Campo Grande, MS, Brazil
}%

\author{ M. Schmidt}
 \email{mateus.schmidt@ufsm.br}
\affiliation{
 Departamento de Física, Universidade Federal de Santa Maria, 97105-900, Santa Maria, RS, Brazil
}%

\begin{abstract}
Motivated by theoretical and experimental studies of the compound ($o$-MePy-V)PF$_6$ reported by Yamaguchi {\it et al.}~\href{https://journals.aps.org/prb/abstract/10.1103/PhysRevB.98.094402}{[Phys. Rev. B 98, 094402]}, we performed a cluster mean-field analysis of an anisotropic Heisenberg model with six competing exchange interactions. 
We study the ground and thermal states by tuning the spin anisotropy, magnetic field, and temperature.
Our results show that an external magnetic field induces quantum fluctuations, suppressing local moments and leading to the occurrence of a magnetization plateau-like state. When a weak spin anisotropy is considered, the competing interactions are affected, and the field-induced fluctuations can lead to a well-defined magnetization plateau within a field range, in which an exotic quantum state can emerge.
This state exhibits the coexistence of ferromagnetic and dimerized chains driven by the relation between frustration and external field.
Moreover, we identify a phase transition from a collinear antiferromagnetic order to a disordered state at a finite temperature.
Our findings reveal unconventional magnetic properties at low temperatures that can guide future experimental studies of verdazyl-based compounds  
with anisotropic exchange interactions.
\end{abstract}

\maketitle

\section{INTRODUCTION}

Spin systems hosting both frustration and quantum fluctuations have proven to be an excellent laboratory for studying emergent phenomena in condensed matter physics. 
The concurrent experimental and theoretical investigations of these systems are fundamental to the advances achieved in the past five decades.
Significant efforts have been devoted to low-dimensional frustrated magnetic systems that provide unique platforms for the onset of complex quantum effects and enhanced quantum fluctuations.
The conflicting situation introduced by frustrated interactions in this class of magnets can lead to exotic states of matter, including quantum spin liquids  \cite{LIU20221034,Morita},  resonating valence bond states \cite{han2023resonating}, collinear and non-collinear antiferromagnets  \cite{WIESER2021168414,soldatov2023low} to name a few. 

Even more interesting phenomena can arise in frustrated antiferromagnets in the presence of an external magnetic field.
An outstanding example is the honeycomb compound $\alpha$-RuCl$_3$, which exhibits a zigzag antiferromagnetic (AF) phase
\cite{PhysRevB.91.144420, banerjee2016proximate}. Under an external magnetic field, this long-range order gives place to a field-induced quantum spin liquid state \cite{yadav2016kitaev}, in which magnetic excitations \cite{banerjee2018excitations} and thermal transport \cite{PhysRevLett.120.117204} are consistent with the spin fractionalization characteristic of the Kitaev model \cite{takagi2019concept}. Numerical results from an extended Kitaev-Heisenberg model with nearest-neighbor couplings reproduce very well the magneto-thermodynamics of $\alpha$-RuCl$_3$ \cite{li2021identification}, suggesting that the anisotropic coupling of spin components is the main source of frustration in this compound.

Another relevant perspective for the onset of magnetic frustration is the mixture of ferromagnetic (FE) and AF nearest-neighbor interactions in bipartite lattices. 
This scenario takes place in the Verdazyl-based salt ($o$-MePy-V)PF$_6$, where $o$-MePy-V = 3-(2-methylpyridyl)-1,5-diphenylverdazyl, synthesized by Yamaguchi and collaborators \cite{Yamaguchi}.  
Crystallographic data indicate that $o$-MePy-V molecules form square lattice layers of $S=1/2$ spins with PF$_6$ anions acting as spacers between layers. 
Experimental measurements show the presence of a sharp peak in the zero-field specific heat, indicating a transition to an AF phase at Néel temperature $T_N \approx 1.7$ K.
The Curie-Weiss temperature $\theta_{CW} \approx -8.2$ K and the frustration parameter $f = \theta_{CW}/T_N \approx 5$ indicate dominant AF interactions with a potential reduction of the ordering temperature driven by frustration.
A fascinating scenario arises when an external magnetic field is applied to the compound ($o$-MePy-V)PF$_6$. 
Contrary to the usual rapid increase and convex form \cite{PhysRevB.95.235135, PhysRevLett.82.3168, PhysRevLett.114.227202}, the low-temperature magnetization as a function of the external field exhibits a gradual increase with alternating concavity. In addition, when magnetization approaches half the saturation value, a plateau-like feature takes place. First-principle molecular orbital calculations indicate the presence of six different exchange interactions between nearest neighbors, which introduce magnetic frustration in the system. In addition, a theoretical analysis using the Heisenberg model and the tensor network method was also carried out in Ref. \cite{Yamaguchi}.
This approach confirmed the existence of a 1/2-plateau-like state and a reduction in average local moments as the magnetic field was increased.
The tensor network is a powerful method to extract the physical properties of the ground state \cite{zhao2010renormalization}.
However, the technique can have limitations when studying frustrated lattices at finite temperatures, particularly in models that undergo phase transitions \cite{poilblanc2021finite}.
In such cases, the correlation length can exhibit exponential growth, especially as the system approaches its critical temperature,
reducing the effectiveness of the technique \cite{Song, Czarnik}.
Therefore, further theoretical approaches are still needed to explore this interesting compound, particularly when considering not only frustration and quantum fluctuations but also anisotropy effects and thermal fluctuations.


The present work aims to investigate the isotropic and anisotropic frustrated spin-1/2 Heisenberg model on a square lattice with nearest-neighbor exchange interactions, motivated by predictions for the compound  ($o$-MePy-V)PF$_{6}$ \cite{Yamaguchi}. 
 By tuning the spin anisotropy, we uncover nontrivial field-induced magnetic phases and elucidate the microscopic mechanisms underlying the plateau-like behavior observed in ($o$-MePy-V)PF$_{6}$. Moreover, we analyze the interplay between field-induced quantum fluctuations and magnetic frustration, demonstrating their critical role in stabilizing magnetization plateaus in anisotropic spin models and in determining the associated thermodynamic responses.
For this study, we employ the cluster mean-field (CMF) method that accurately captures the ground state and enables the estimation of finite-temperature effects on thermodynamic quantities. 
It has been successfully applied to study phase transitions in various magnetic systems, including the Ising model on square \cite{jin2013phase,godoy2020ising}, triangular \cite{malakar2020phases}, and kagomé lattices \cite{schmidt2017spin}, as well as the Ising model with quantum fluctuations \cite{kellermann2019quantum,zimmer2016quantum} and the Heisenberg model with competitive interactions \cite{WIESER2021168414}. 
Additionally, the CMF approach has been used to study frustrated systems under magnetic fields, qualitatively reproducing experimental results for compounds such as ${\mathrm{CuInVO}}_{5}$ \cite{Singhania} and 
$\mathrm{Tm}\mathrm{Mg}\mathrm{Ga}{\mathrm{O}}_{4}$ \cite{PhysRevResearch.2.043013}.

The remainder of the paper is organized as follows. In Sec. \hyperref[sec:model]{II}, we present the model and the CMF framework. In Sec. \hyperref[sec:results]{III}, we present results for the ground state at zero temperature and discuss the roles of thermal and enhanced quantum fluctuations on the phase diagram of the model. Finally, Sec. \hyperref[sec:summary]{IV} summarizes the paper and presents our conclusions.
\begin{figure}[t!]
    \centering    \includegraphics[width=0.75\columnwidth] {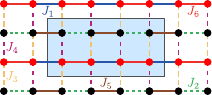}
   \caption{ Schematic representation of the anisotropic square lattice with six different exchange couplings $J_{n}$ (\{$n=1,...,6\}$). Solid (dashed) lines denote FE (AF) exchange interactions. The red sites form a FE chain with $J_1$ and $J_6$ exchange couplings (referred to as chain 1). The black sites form a chain with AF and FE exchange couplings $J_{2}$ and $J_{5}$ (referred to as chain 2). These chains are coupled by AF interactions $J_{3}$ and $J_{4}$, resulting in a frustrated square lattice. Square loops containing an odd number of AF interactions are frustrated. The unit cell of the compound is highlighted in blue color.}
    \label{fig:square-lattice}
\end{figure}
\section{MODEL AND METHOD}\label{sec:model}
In this work, we consider a version of the anisotropic Heisenberg model on the square lattice with six different constant couplings $J_{n}$ ($\{n=1,...,6\}$).
The Hamiltonian is given by
\begin{equation}\label{eq:ham}
    H= -\sum_{\langle i,j\rangle} J_{n}[(1-\Delta)(\sigma_i^x \sigma_j^x+\sigma_i^y \sigma_j^y) + \sigma_i^z \sigma_j^z] - h^{z}\sum_{i} \sigma^z_i,
\end{equation}
where $\sigma_{i}^{\alpha}$ are the spin-1/2 Pauli operators with $\alpha=\{x,y,z\}$ and $J_{n}$ represents the anisotropic exchange couplings between nearest-neighbor sites $i$ and $j$. 
The last term of Eq. (\ref{eq:ham}) incorporates the magnetic field $h^{z}$.
The anisotropy parameter $\Delta$ allows recovering different limits of the Hamiltonian (\ref{eq:ham}) in the range $0\le \Delta \le 1$ ($\Delta=0$ corresponds to the isotropic Heisenberg model and $\Delta=1$ corresponds to the Ising limit). 
In this easy-axis anisotropic system, tuning $\Delta$ allows going from a scenario in which strong quantum fluctuations are expected ($\Delta=0$) to a classical model.

The set of couplings $J_{n}$ follows that presented in Ref. \cite{Yamaguchi} and is depicted in Fig. \ref{fig:square-lattice}.
The red sites (chain 1) form a linear chain of quantum spins coupled by FE exchange interactions $J_1$ and $J_6$.
On the other hand, the black sites (chain 2) are coupled by the interactions $J_2<0$ and $J_5>0$, resulting in an AF-FE quantum spin chain. 
These two types of quantum spin chains are coupled by the AF interactions $J_3$ and $J_4$, producing a 2D frustrated square lattice. 
It is worth noting that several square loops in the lattice contain an odd number of AF interactions, indicating frustration. Alternatively, a frustrated loop can be identified by a negative value of the product of exchange interactions within the loop, i.e., $\Pi_{n } J_n<0$. We note that all loops containing the exchange couplings $J_2$ and $J_6$ are frustrated.

\begin{figure}[!t]
    \centering
    \includegraphics[width=.36\textwidth]{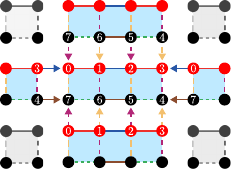}
    \caption{Schematic representation of eight-site cluster mean-field method. The arrows represent couplings approximated within the mean-field level. Interactions were represented following the same convention of Fig. \ref{fig:square-lattice}.}
    \label{fig:CMF}
\end{figure}

We investigate the frustrated model defined in Eq. (\ref{eq:ham}) employing the CMF theory, where the first step is to divide the infinite lattice into $N_{cl}$ identical clusters of $n_s$ sites.
In particular, we divided this lattice into clusters with eight sites, following the unit cell of compound ($o$-MePy-V)PF$_{6}$ depicted in Fig.  \ref{fig:square-lattice}.  The choice of cluster in the CMF scheme is illustrated in Fig.  \ref{fig:CMF}.
In this CMF approach, we can write the Hamiltonian as 
\begin{equation}\label{eq:hcmf1}
H=\sum_{\mu=1}^{N_{cl}}\left(H_{\mu} - h^{z}\sum_{i \in \mu}\sigma^z_{i}\right)+\sum_{\langle\mu,\nu\rangle}H_{\mu\nu},
\end{equation}
where the exact contribution of intracluster interactions of cluster $\mu$ is given by 
\begin{equation}
H_{\mu} = -\sum_{\langle i,j\rangle_{\mu}} J_{n} [\sigma_{i}^z \sigma_{j}^z +(1-\Delta)(\sigma_{i}^x \sigma_{j}^x+\sigma_{i}^y \sigma_{j}^y)] ,
\end{equation}
where $\langle i,j\rangle_{\mu}$ denotes a sum over the nearest neighbor sites that belong to cluster $\mu$. The intercluster contribution is represented by
\begin{equation}
H_{\mu\nu} = -\sum_{\langle i,j\rangle_{\mu\nu}} J_{n} [\sigma_{i}^z  \sigma_{j}^{z} 
+(1-\Delta)(\sigma_{i}^x  \sigma_{j}^x + 
\sigma_{i}^y  \sigma^{y}_{j})],
\label{eq:inter}
\end{equation}
where $\langle i,j\rangle_{\mu\nu}$ denotes a sum over nearest neighbor sites $i$ and $j$ that belong to different clusters $\mu$ and $\nu$.

In our CMF scheme, the intercluster interactions are treated by the mean-field approximation 
\begin{equation}\label{eq:mean}
    \sigma_{i}^\alpha \sigma_{j}^\alpha \approx \sigma_{i}^\alpha  m^{\alpha}_{j} + m^{\alpha}_{i}\sigma_{j}^\alpha - m^{\alpha}_{i}m^{\alpha}_{j},
\end{equation}
for each component $\alpha=x$, $y$ and $z$, in which $m^{\alpha}_{i} = \langle \sigma^{\alpha}_{i} \rangle$ is the $\alpha$ component of the local magnetization of site $i$,  with $\langle \cdots \rangle$ denoting the Boltzmann weighted average.
In the next section,  $m^{z}_{i}$ and $m^{x}_{i}$ will be referred to as the longitudinal ($m^{||}_{i}$) and transverse ($m^{\bot}_{i}$) local magnetizations relative to the external magnetic field, respectively. It is important to emphasize that the transitions occur within the $xz$ plane, where $m^{y}_{j}=\langle \sigma^{y}_{j} \rangle=0$.

Furthermore, we assume that the spatial translation symmetry of the clusters leads to an identical pattern of local magnetizations in all clusters. Therefore, instead of evaluating the local magnetizations for all sites in the system, the problem is reduced to computing the set of local magnetizations for a particular cluster.  
 This allows us to write a self-consistent effective single-cluster problem as
\begin{equation}\begin{split}\label{eq:hcmf}
H_{\textrm{eff}} =& -\sum_{\langle i,j\rangle} J_{n} [\sigma_{i}^z \sigma_{j}^z +(1-\Delta)(\sigma_{i}^x \sigma_{j}^x+\sigma_{i}^y \sigma_{j}^y)] 
\\
&-\sum_{(i,k)}J_{\bar{n}}[\sigma_{i}^z  m^{z}_{k} +(1-\Delta)(\sigma_{i}^x  m^{x}_{k}
+\sigma_{i}^y  m^y_{k} \\ &-m^{x}_{i}m^{x}_{k}/2-m^{y}_{i}m^{y}_{k}/2) -m^{z}_{i}m^{z}_{k}/2]
 \\ & - h^{z}\sum_{i}\sigma^z_{i}, 
\end{split}\end{equation}
where all sums run over sites of a single cluster.
The sum of indices $(i,k)$ runs for all edge sites $i$ of the central cluster, with the corresponding exchange interaction $J_{\bar{n}}$ between sites $i$ and its neighbor cluster.  
 For example, when $i=0$, following Fig. \ref{fig:CMF}, we have $k=3$ with $J_{\bar{n}}=J_1$ and $k=7$ with $J_{\bar{n}}=J_4$.
 Therefore, the self-consistent local magnetizations  of the effective model are evaluated from
 \begin{equation}\label{eq:local_mag}
     m_{i}^{\alpha}= \frac{\textrm{Tr} \, \sigma_{i}^{\alpha}\, e^{-\beta H_{\textrm{eff}}}}{Z},
 \end{equation}
where $Z= \textrm{Tr} \, e^{-\beta H_{\textrm{eff}}}$ is the canonical partition function and $\beta=1/(k_B T)$ ($k_B$ is the Boltzmann constant and $T$ is temperature).
After finding the self-consistent solution for the local magnetizations, we can compute the thermodynamic quantities for the system. For instance, the total magnetization is given by
\begin{equation}
M=\left[{\left(M^{x}\right)}^2+{\left(M^{y}\right)}^2+{\left(M^{z}\right)}^2\right]^{1/2},
\end{equation}
where $M^{\alpha}=\sum_{j}m_{j}^{\alpha}/n_s$, with the sum performed over all sites from the cluster. We can also compute the spin-spin correlation functions from
\begin{equation}
    C_{ij} = \braket{\sigma_{i}^{x}\sigma_{j}^{x}+\sigma_{i}^{y}\sigma_{j}^{y}+\sigma_{i}^{z}\sigma_{j}^{z}},
\end{equation}
the free energy per spin
\begin{equation}
        f = -\frac{k_B T}{n_{s}}\ln{(Z)},
\end{equation}
and the enthalpy per spin
\begin{equation}
    u = \frac{\textrm{Tr}\,\{H_{\textrm{eff}}\,e^{-\beta H_{\textrm{eff}}}\}}{n_s Z}.
\end{equation}
The magnetic entropy per spin can be obtained from $S = (u - f)/T$ and the specific heat can be evaluated from the enthalpy per spin by computing
\begin{equation}
     C_{h^{z}} = \left(\frac{\partial u}{\partial T}\right)_{h^{z}}.
\end{equation}

 It is important to remark that these thermodynamic quantities can reveal important insights into the physics of the model system. The frustrated systems in the presence of magnetic fields and thermal fluctuations can present novel phase transitions, or sometimes, multiple phase transitions \cite{PhysRevB.101.064403}. 
These phase transitions are manifested as anomalies or discontinuities in thermodynamic quantities, such as specific heat, spin-spin correlations, magnetic susceptibility, and local magnetizations.
The free energy is used to choose the thermodynamically stable state \cite{ROOS2024129979}.

\section{RESULTS AND DISCUSSION}\label{sec:results}

We obtain numerical results by solving the set of local magnetizations given by Eqs. (\ref{eq:local_mag}) and (\ref{eq:hcmf}) self-consistently, in which a finite cluster of up to 8 sites is adopted and exactly diagonalized. These magnetizations enable us to obtain an upper bound of the free energy, from which we derive the other thermodynamic quantities. 
For numerical purposes, we adopt $k_B$ as unity. 
The FE exchange interaction $J_1$ is used as a unit of energy. Other exchange couplings follow the relations $J_2/J_1=-0.82$, $J_3/J_1=-0.66$, $J_4/J_1=-0.61$, $J_5/J_1=0.26$ and $J_6/J_1=0.20$, as suggested by Ref. \cite{Yamaguchi}. The magnetic properties of the ground state are first analyzed by setting $T/J_1=0$. After, we present the effects of thermal fluctuations considering the dependence of temperature on the quantities.

\subsection{Properties of the ground state}

\begin{figure}[!t]
    \centering    \includegraphics[width=.45\textwidth]{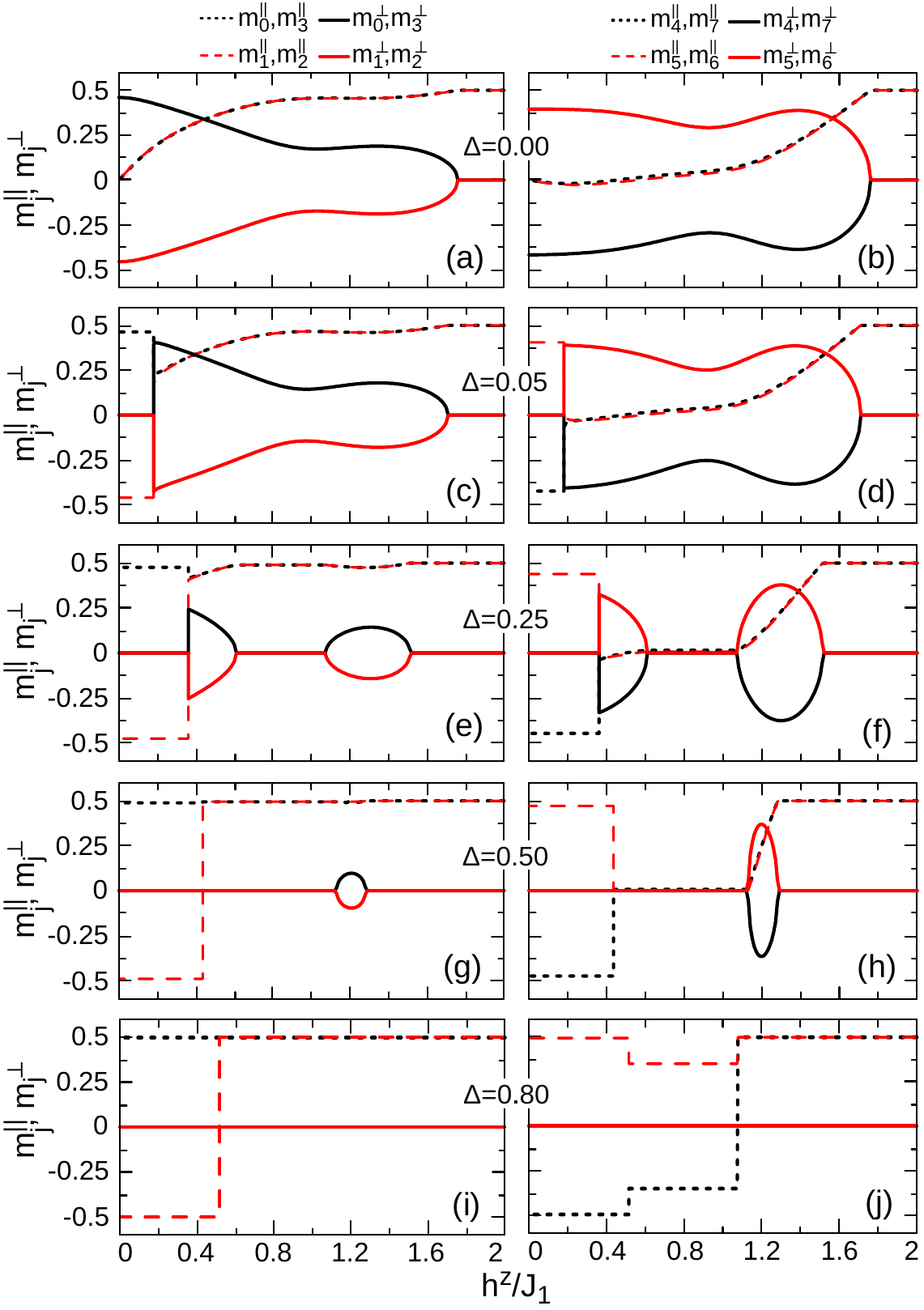}
    \caption{Longitudinal (dashed lines) and transverse (solid lines) components of the eight local magnetizations in the ground state under an external magnetic field  and anisotropy values (a)\label{meanfields(a)} and (b)\label{meanfields(b)} $\Delta=0.0$, (c)\label{meanfields(c)} and (d)\label{meanfields(d)} $\Delta=0.05$, (e)\label{meanfields(e)} and (f)\label{meanfields(f)} $\Delta=0.25$, (g)\label{meanfields(g)} and (h)\label{meanfields(h)} $\Delta=0.50$ and (i)\label{meanfields(i)} and (j)\label{meanfields(j)} $\Delta=0.80$. The left column displays the local moments of chain 1, while the right column shows the local moments of chain 2.
       }
        \label{fig:meanfields2}
\end{figure} 

A notable feature of ($o$-MePy-V)PF$_6$ is its molecular structure, which consists of two quantum spin chains coupled by AF interactions \cite{Yamaguchi}.
The longitudinal ($m^{||}_{j}$) and transverse ($m^{\bot}_{j}$) magnetization components of these chains are shown in Fig. \ref{fig:meanfields2} as a function of the applied magnetic field for selected values of the anisotropic parameters. 
For the isotropic Heisenberg system, the longitudinal magnetizations are null for zero magnetic fields, whereas the transverse magnetizations are finite and antisymmetric.  
When the magnetic field is turned on,
the longitudinal components increase monotonically until $h^{z}/J_1\approx 1.00$, become roughly constant, and then saturated at $h^{z}/J_1\ge 1.75$ (see the dotted and dashed lines in panel \hyperref[meanfields(a)]{(a)}).
In contrast, the longitudinal components of chain 2 exhibit a slight but unusual decrease in weak magnetic fields 
(see the dotted and dashed lines in panel \hyperref[meanfields(b)]{(b)}). 
This decrease can be an effect of quantum fluctuations induced by the field, and it is qualitatively in agreement with the findings reported in Ref. \cite{Yamaguchi}.
On the other hand, the intensity of the transverse components of both chains decreases with the magnetic field until $h^{z}/J_1\approx 1.00$, where an increase can be observed before they become null in the critical field.

The local magnetization of two chains suggests a collinear AF order with a twofold periodicity structure in the transverse plane, where $m_{1}^{\bot}\approx m_2^{\bot}$ and $m_{4}^{\bot}\approx m_7^{\bot}$ are negative and $m_{0}^{\bot}\approx m_3^{\bot}$ and $m_{5}^{\bot}\approx m_6^{\bot}$ are positive. 
Although the interactions could suggest an FE ordering in chain 1, this collinear AF order arises due to the interchain AF interactions $J_{3}$ and $J_{4}$. 
In this case, frustration and quantum fluctuations, enhanced by the applied field, can play significant roles alongside the gradual suppression of collinear order by the Zeeman effect.

When a weak anisotropy is considered ($\Delta=0.05$ in panels \hyperref[meanfields(c)]{(c)} and \hyperref[meanfields(d)]{(d)} of Fig. \ref{fig:meanfields2}), the local magnetizations exhibit a different behavior at low field intensities.
The longitudinal components are finite and antisymmetric, remaining independent of the field, while the transverse components are null in the range $0 \leq h^z/J_1 \lesssim 0.2$. 
This independence of the local magnetizations reflects on the system properties, resulting in a zero magnetization plateau for weak fields.
A collinear AF order is also found in the presence of anisotropy, but it is oriented along the field direction.
This AF phase undergoes a spin-flop transition at $h^z/J_1 \approx 0.2$. 
In contrast to the case $h^{z}/J_{1} = 0$, where the self-consistent set of equations for the local magnetizations converges to a unique solution, for $h^{z}/J_{1} > 0$ there exists a coexistence of solutions within a certain range of magnetic field, each corresponding to distinct local magnetization patterns. In this situation, the free energies of the different solutions are compared to determine the thermodynamically stable phase and to identify the transition point \cite{keffer1973dynamics,bogdanov2007spin,machado2017spin}.
For $h^z/J_1 > 0.2$, similar results to the isotropic model with a unique solution are observed. 

The regimes for $\Delta = 0.25$ (panels \hyperref[meanfields(e)]{(e)} and \hyperref[meanfields(f)]{(f)} of Fig. \ref{fig:meanfields2}) and $\Delta = 0.50$ (panels \hyperref[meanfields(g)]{(g)} and \hyperref[meanfields(h)]{(h)} of Fig. \ref{fig:meanfields2}) exhibit similar results for the local magnetizations under a weak applied field, where the longitudinal components are finite while the transverse components vanish.
However, as the field increases, the transverse components can become finite, indicating that field-induced quantum fluctuations can drive spin-flop transitions in these anisotropic systems. In both anisotropic regimes, the local moments of chain 2 display unusual magnetic behavior, with all components decreasing to zero (see the range $0.6<h^{z}/J_{1}<1.1$ for $\Delta = 0.25$ in Fig. \ref{fig:meanfields2}\hyperref[meanfields(f)]{(f)} and $0.4<h^{z}/J_{1}<1.1$ for $\Delta = 0.50$ in Fig. \ref{fig:meanfields2}\hyperref[meanfields(g)]{(g)}).
As we will discuss, this behavior is associated with a magnetization plateau state and reveals intriguing physics that may be linked to an exotic state of matter.
The increase in anisotropy for $\Delta=0.80$ (panels \hyperref[meanfields(i)]{(i)} and \hyperref[meanfields(j)]{(j)} of Fig. \ref{fig:meanfields2}) is mainly characterized by the absence of transverse components in the local moments, indicating a classical regime.

\begin{figure}[!t]
    \centering
    \includegraphics[width=0.49\textwidth]{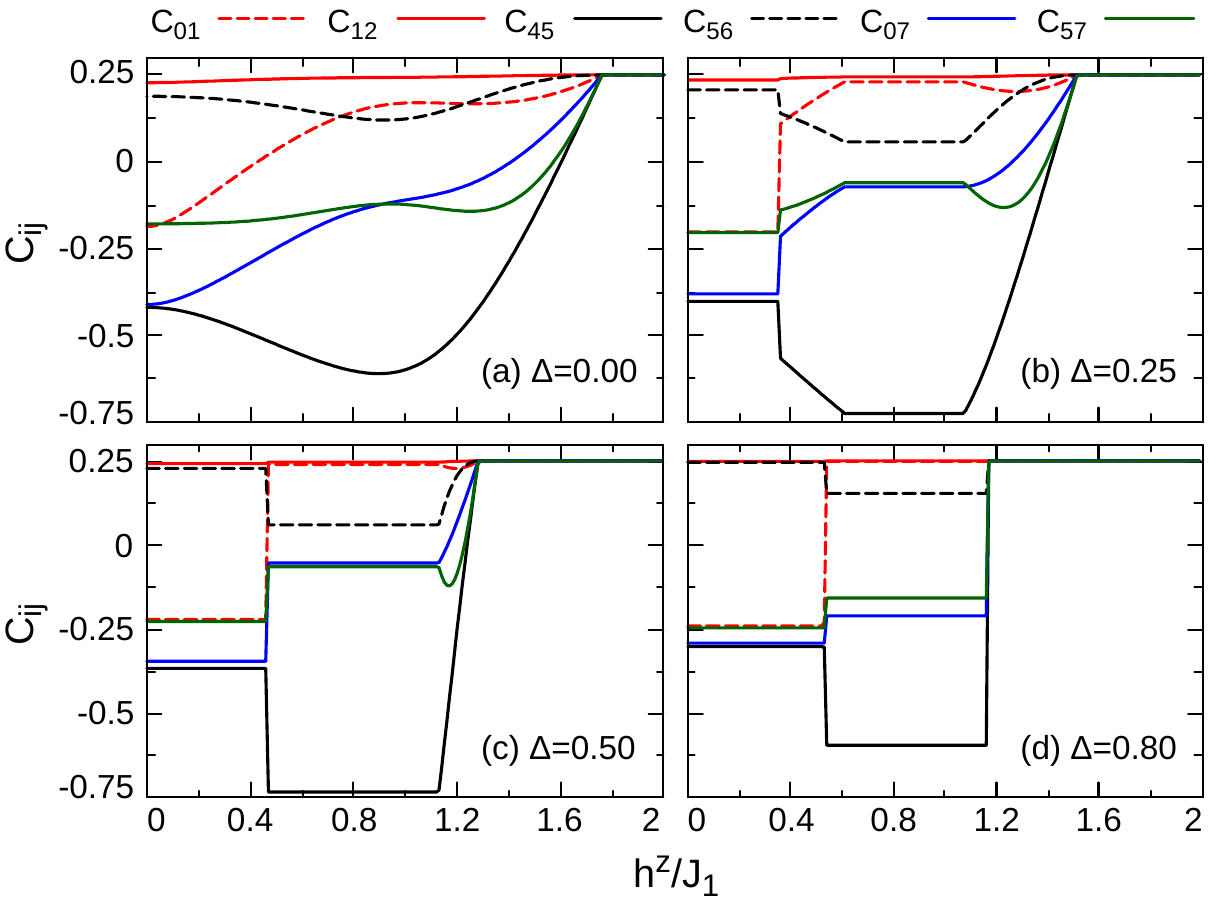}
    \caption{Field dependence of spin-spin correlation functions computed using 8-site CMF at $T/J_{1}$=0 and anisotropic values (a) $\Delta=0.00$\label{fig:c_delta0}, (b) $\Delta=0.25$\label{fig:c_delta025}, (c) $\Delta=0.50$\label{fig:c_delta050}, (d) $\Delta=0.80$\label{fig:c_delta080}. Some correlations have been omitted from the figure as they are nearly equivalent to those already shown, e.g., the correlations $C_{01}\approx C_{23}$, $C_{45}\approx C_{67}$, $C_{46}\approx C_{57}$ and the interchain correlations $C_{07}\approx C_{16} \approx C_{25} \approx C_{34}$. }
    \label{fig:correlations}
\end{figure}

The spin-spin correlation functions $C_{ij}$ can help characterize the ground state  in different anisotropic scenarios.
For example, Fig. \ref{fig:correlations}\hyperref[fig:c_delta0]{(a)} shows the field dependence of the $C_{ij}$ 
for the isotropic model. 
At zero magnetic field, the $C_{ij}$ 
for chains 1 and 2 are consistent with a collinear AF state, where $C_{01}$, $C_{23}$, $C_{45}$ and $C_{67}$ exhibit AF behavior, while $C_{12}$ and $C_{56}$ show FE correlations. 
As the magnetic field increases, the correlations $C_{45}$ and $C_{67}$ exhibit an enhanced AF character, which can be attributed to  the increased quantum fluctuations.
This effect can also be seen for the anisotropy $\Delta=0.25$ in Fig. \ref{fig:correlations}\hyperref[fig:c_delta025]{(b)} and $\Delta=0.50$ in Fig. \ref{fig:correlations}\hyperref[fig:c_delta050]{(c)}.
In these cases, the correlations $C_{45}$ and $C_{67}$ can assume the perfect singlet value of \textminus0.75, while the interchain correlations (e.g. $C_{07}$, $C_{16}$,...) and $C_{56}$ become very small. 
Remarkably, this suggests that the ground state is effectively characterized by FE ordering in chain 1 and the formation of singlet pairs in chain 2. 
In addition, besides $C_{56}$, the correlations $C_{46}$ and $C_{57}$ in chain 2 are also small, indicating that this chain effectively forms singlet pairs and becomes dimerized.
Notably, the interplay between frustration and enhanced quantum fluctuations appears more pronounced over this range of anisotropy.
For $\Delta=0.80$ in Fig. \ref{fig:correlations}\hyperref[fig:c_delta080]{(d)}, the previously described singlet pair formation is absent, but chain 2 still maintains a collinear spin alignment before saturation.

\begin{figure}[!t]
    \centering
    \includegraphics[width=.45\textwidth]{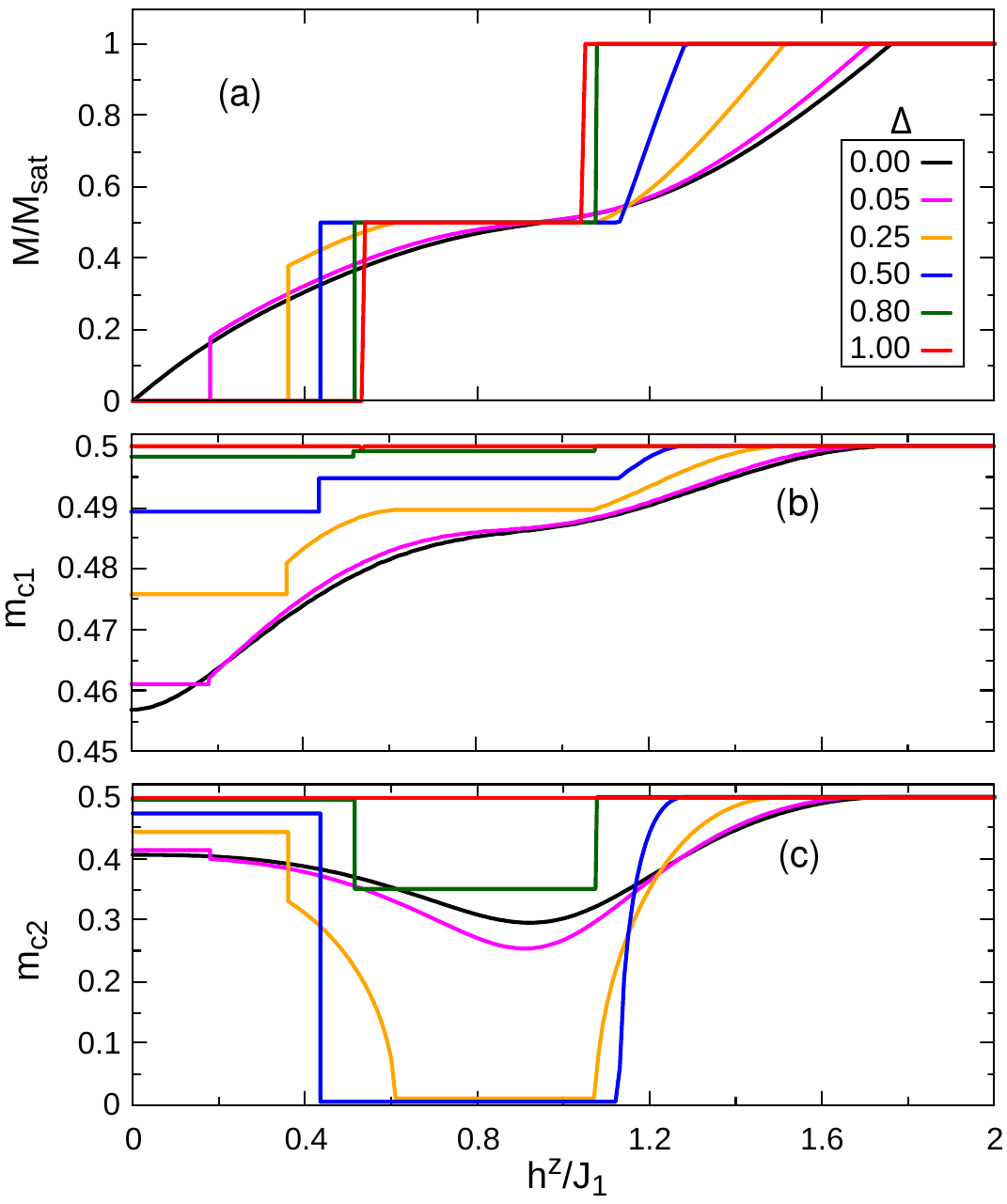}
    \caption{Ground state of the CMF model under magnetic field and different anisotropy values. (a)\label{fig:magt} Total magnetization curve normalized by the saturation, (b)\label{fig:mc1} average local moments of chain 1 $m_{c1}$ and (c)\label{fig:mc2}  average local moments of chain 2 $m_{c2}$.}
    \label{fig:mags}
\end{figure}

Figure \ref{fig:mags}\hyperref[fig:magt]{(a)} shows the field dependence of total magnetization for different anisotropy parameters.
In the Ising limit ($\Delta=1.00$), a zero magnetization plateau is observed, characterized by the collinear AF order in the longitudinal direction. For $0.5\lesssim h^z/J_1\lesssim 1.0$, a plateau appears at exactly one-half of the saturation, with chain 1 fully polarized in the field direction, while chain 2 preserves a collinear AF order.
In this case, the system undergoes a spin-flip transition to a state with magnetization $M/M_{sat}=1/2$ before reaching saturation. 
   
For $0.25\le\Delta \le 0.80$, the zero magnetization plateau remains consistent with previous descriptions. 
However, increasing $\Delta$ leads to subtle changes in magnetization curves, such as smaller jumps and a gradual increase near the saturation field.
To clarify these features, we introduce the average local moments for both chains in  Eqs. (\ref{eq:mc1}) and (\ref{eq:mc2}) as:
\begin{equation}\label{eq:mc1}
    m_{c1}^{2}=\frac{1}{4}\sum_{j=0}^{3} ({\langle \sigma_j^x\rangle}^2+{\langle \sigma_j^y\rangle}^2+{\langle \sigma_j^z\rangle}^2),
\end{equation}
and
\begin{equation}\label{eq:mc2}
    m_{c2}^{2}=\frac{1}{4}\sum_{j=4}^{7} ({\langle \sigma_j^x\rangle}^2+{\langle \sigma_j^y\rangle}^2+{\langle \sigma_j^z\rangle}^2).
\end{equation}
For instance, the average local moments of chain 1 gradually decrease as $\Delta$ approaches the isotropic limit, but remain near their maximum value of 1/2 over the entire field range (see Fig. \ref{fig:mags}\hyperref[fig:mc1]{(b) }for $\Delta = 0.80$, $\Delta = 0.50$, and $\Delta = 0.25$). 
Furthermore, $m_{c1}$ increases with field intensity but presents a constant value within the plateau range.
In contrast, chain 2 undergoes a significant reduction in the average local moment due to the applied field (see Fig. \ref{fig:mags}\hyperref[fig:mc2]{(b)}). 
For $\Delta = 0.50$ and $\Delta = 0.25$, $m_{c2}$ approaches zero in the 1/2-plateau region, indicating
that frustration and enhanced quantum fluctuations destroy local magnetic order. This drives the chain into a nonmagnetic phase with $\braket{\sigma_{i}^{x}}\approx\braket{\sigma_{i}^{z}}\approx0$, while spin-spin correlations exhibit clear dimerization. 
The shift from 1/2-plateau to saturation is characterized by a smooth magnetization increase, followed by the emergence of transverse components (see Fig. \ref{fig:mags}\hyperref[fig:magt]{(a)}). 
For $\Delta = 0.25$, this transition also occurs in weak fields, where chain 2 evolves from a collinear AF order in the longitudinal direction to a transverse one, followed by a transition to a dimerized phase before becoming fully polarized at higher fields.
In this process, chain 1 remains polarized while exhibiting a small transverse contribution (see Fig. \ref{fig:meanfields2}\hyperref[meanfields(e)]{(e)}).
     
Reducing the anisotropy to $\Delta = 0.05$ and $\Delta = 0.00$ reveals new insights. 
In both cases, the magnetization exhibits a plateau-like behavior rather than a well-defined magnetic plateau. Even weak anisotropy induces a zero-magnetization plateau, followed by a spin-flop transition to finite magnetization as the field increases (as discussed in Fig. \ref{fig:meanfields2}).
The average local moments of chain 1 follow the previous description, being less affected by the external field.  Conversely, chain 2 exhibits a reduction in average local moments, though, unlike in higher anisotropy cases ($\Delta=0.25$ and $0.50$), they remain finite. Furthermore, these moments can reflect the collinear AF order in the transverse direction within the plateau-like region, as illustrated in Fig. \ref{fig:meanfields2}.

These results highlight key aspects:
(i) Chain 1 consistently exhibits strong polarization under varying magnetic fields in different anisotropies, suggesting that it alone does not significantly modify the ground state of the system.
(ii)  Chain 2 is strongly affected by the applied field and anisotropy, exhibiting field-induced quantum fluctuations that can lead to singlet pair formation and nonmagnetic states.    
This high sensitivity to $h^{z}$ and $\Delta$ is likely a consequence of the alternating FE and AF exchange interactions within this quantum spin chain.
(iii) The AF coupling between the chains introduces competitive interactions, leading to frustration. Near the Ising limit, this frustration stabilizes a one-half magnetization plateau. In contrast, near the isotropic limit, quantum fluctuations drive the system toward a plateau-like state rather than a well-defined plateau.

\begin{figure}[!t]
    \centering
    \includegraphics[width=.45\textwidth]{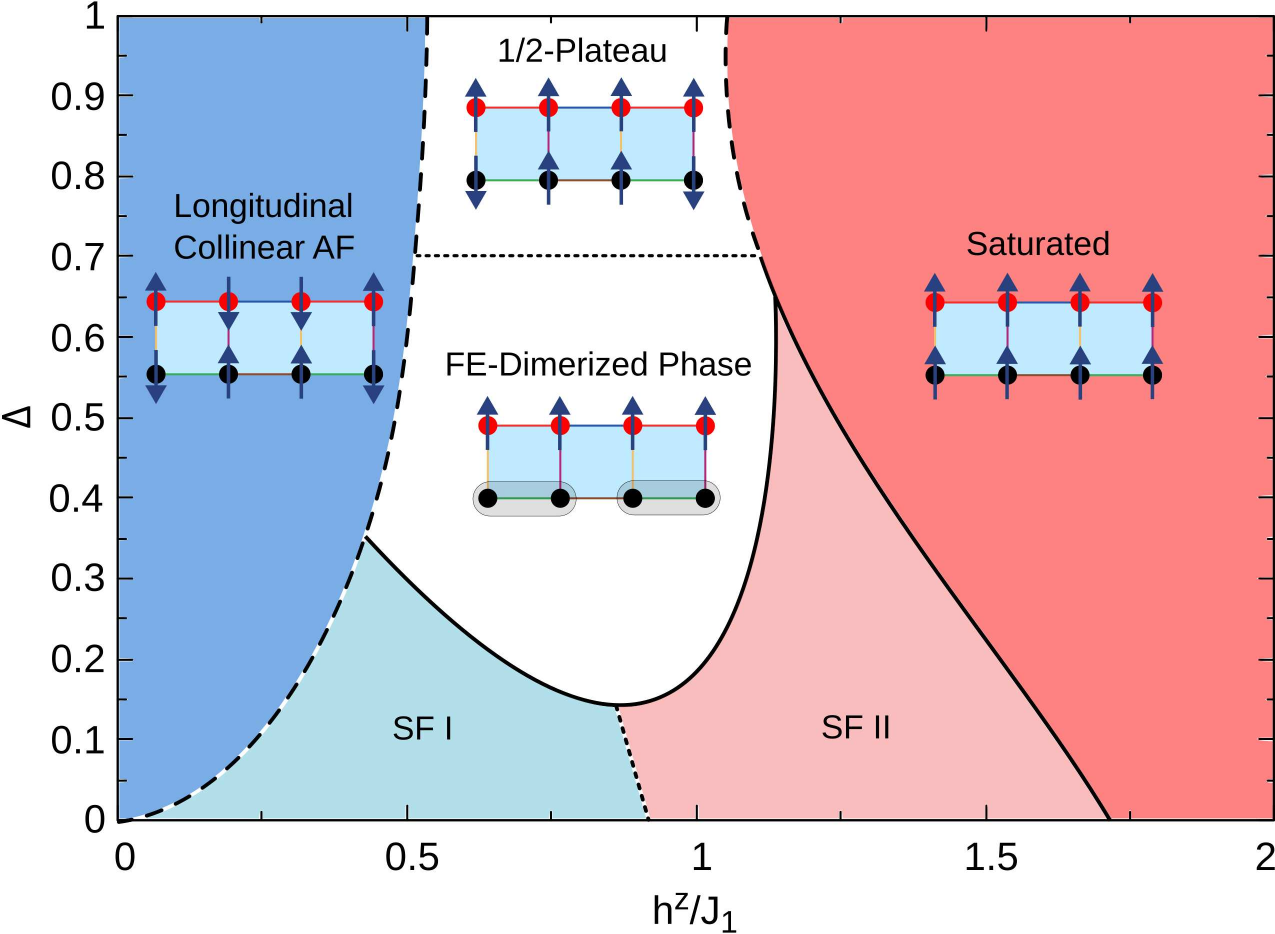}
    \caption{ Ground-state phase diagram obtained from the CMF analysis of the anisotropic Heisenberg model on a square lattice with six different interactions. The diagram shows the collinear AF phase at low fields, a one-half magnetization plateau (white region) with classical and FE-dimerized phases, and two spin-flop (SF I and SF II) regimes. At high fields, the system saturates to a fully polarized state. The solid and dashed lines denote second-order and discontinuous phase transitions, respectively, while the dotted lines represent a crossover region between two regimes.}
    \label{fig:phase-diagram}
\end{figure}

The ground-state phase diagram depicted in Fig. \ref{fig:phase-diagram} summarizes the findings of this section. 
As the spin anisotropy increases, the longitudinal collinear AF order dominates at low magnetic fields (see local magnetizations of Fig. \ref{fig:meanfields2}).
At intermediate fields, a one-half magnetization plateau phase emerges (see the white region on the phase diagram), characterized by a classical plateau state at higher values of $\Delta$. 
For intermediate intensities of $\Delta$, the singlet correlation functions (Fig. \ref{fig:correlations}) and the average local moments (Fig. \ref{fig:mags}) support an FE-dimerized phase (delimited by the dotted line in the phase diagram), suggesting a strong interplay of frustration and field-induced quantum fluctuations.
An inflection in the magnetization curve as a function of applied field occurs for a lower spin anisotropy, indicating a plateau-like regime (see Fig. \ref{fig:mags}). 
In this regime, increasing the magnetic field leads to distinct spin-flop transitions (SF I and SF II), which are separated by the inflection point.  
At high fields, the system undergoes a transition to a fully saturated phase. 
In addition, the transitions from the longitudinal collinear AF state to other phases are discontinuous, as is the transition from the classical 1/2-plateau to the saturated state.
These results highlight the role of spin anisotropy in shaping the ground-state.

\subsection{Finite temperature properties}

\begin{figure}[!t]
    \centering
    \includegraphics[width=.45\textwidth]{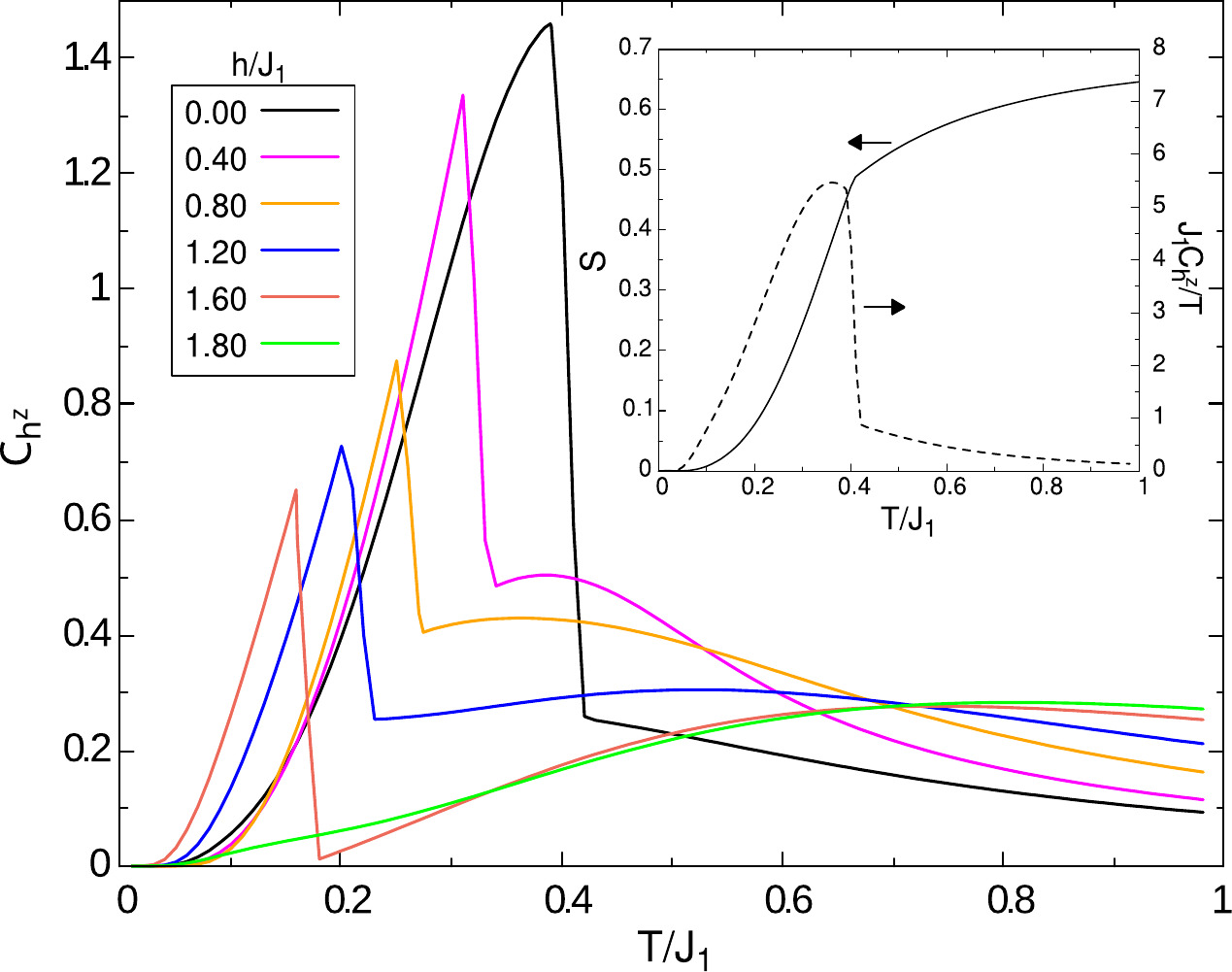}
    \caption{Specific heat as a function of temperature within the 8-site CMF approximation for different magnetic field values and anisotropy $\Delta=0$.  The solid and dashed lines in the inset illustrate the temperature dependence of magnetic entropy (left axis) and specific heat $J_1C_{h^{z}}/T$ (right axis) at zero applied field, respectively. 
    }
    \label{fig:especific-heat}
\end{figure} 

In this section, we investigate the effects of thermal fluctuations on the isotropic spin model.
We explore the specific heat as a function of temperature in Fig. \ref{fig:especific-heat}. 
At zero magnetic field, the specific heat exhibits a sharp peak at $T_{N}/J_{1}=0.40$, which is associated with a long-range order phase transition.
As we increase the magnetic field, the peak temperature decreases until $h^z/J_1\leq 1.6$.
For $h^z/J_{1} > 1.6$, the transition temperature completely disappears due to saturation, without a characteristic peak in the specific heat curve.
It is also worth noting the presence of a round maximum in the magnetic specific heat above the ordering temperature when finite external fields are considered. This feature is often found in frustrated magnetic systems \cite{nath2008magnetic,tutsch2019specific}.
We further observe that the CMF method qualitatively reproduces the experimental findings reported for the compound ($o$-MePy-V)PF$_6$, as detailed in Ref. \cite{Yamaguchi}.
In addition, the local magnetizations exhibit continuous behavior, characterized by a single self-consistent solution across the entire temperature range analyzed for different field intensities, suggesting the occurrence of a single finite-temperature phase transition.

The inset of Fig. \ref{fig:especific-heat} shows the temperature dependence of the magnetic entropy and the specific heat over temperature per spin at zero external magnetic field.
The low-temperature specific heat follows a $T^2$ dependence, in good agreement with experimental data (see the Fig. 2(a) of Ref. \cite{Yamaguchi}). 
This behavior is consistent with the presence of a linearly dispersive mode in a 2D AF system, as predicted theoretically in Ref. \cite{chubukov1994theory}.
The entropy gradually decreases from ln(2) on cooling, reaching zero as $T\rightarrow 0$, consistent with an ordered ground state. 
It is important to remark that although frustration and quantum fluctuations become relevant at lower thermal fluctuations, short-range correlations can still introduce an ordered state at lower temperatures. 
In addition, the release of entropy in the disordered state can be measured by $S_{rel}\equiv \ln(2)-S(T_{N}/J_1)=0.24$ (in $k_B$ units), which is consistent with the reported in Ref. \cite{Yamaguchi}.
As expected, the sharp peak in specific heat coincides with the point of concavity change in magnetic entropy, suggesting that the system indeed undergoes a second-order phase transition at $T_{N}$.
We emphasize that similar behavior has also been observed in other verdazyl-based charge-transfer salts, such as ${\mathrm{[Zn(hfac)_{2}](4-Br-o-Py-V)}}$ \cite{Yamaguchi2021} and ${\mathrm{(o-MePy-V)FeCl_{4}}}$ \cite{Ywasaki}.

\begin{figure}[!t]
    \centering
\includegraphics[width=.42\textwidth]{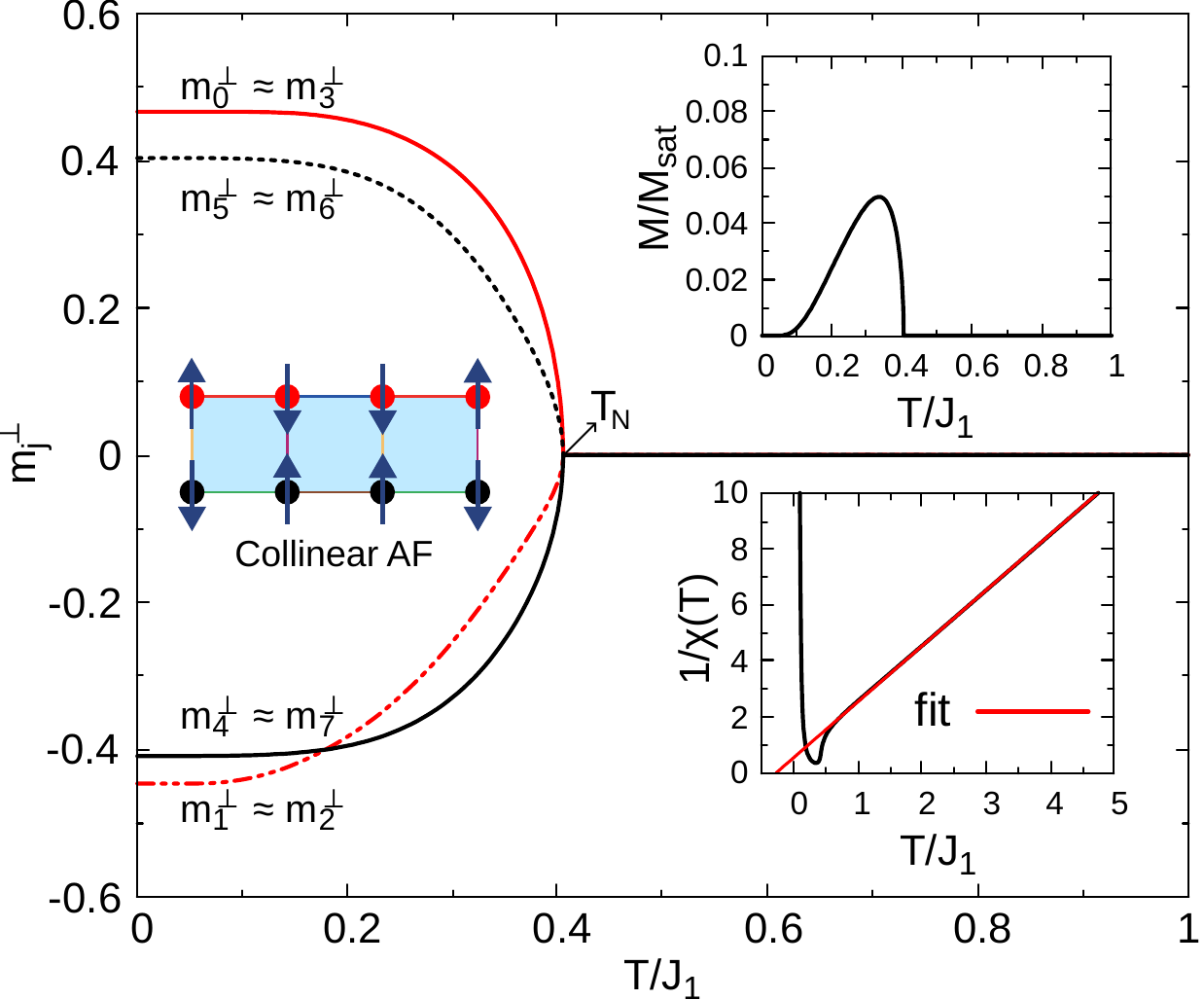}
     \caption{
    The temperature dependence of local magnetizations in the absence of an external magnetic field is shown. In our chosen axes, the local magnetization components in the $z$-direction are null, while the transverse components are displayed in the figure plane.
    The bottom inset illustrates the inverse susceptibility $1/\chi(T)$ and the top inset shows the contribution to the temperature dependence of total magnetization, reflecting a small asymmetry of local magnetizations. We also provide insets with schematic representations of the collinear phase.}
    \label{fig:mag}
\end{figure}
The second-order phase transition can be corroborated by local magnetization curves in Fig. \ref{fig:mag}, where a clear transition from the high-temperature disordered phase to an ordered phase with nonzero local magnetizations occurs at $T_N$. 
As discussed in Figs. \ref{fig:meanfields2}, there are four sets of local magnetizations in the ground state $m_{0}^{\bot}\approx m_{3}^{\bot}$, $m_{1}^{\bot}\approx m_{2}^{\bot}$, $m_{4}^{\bot} \approx m_{7}^{\bot}$, and $m_{5}^{\bot}\approx m_{6}^{\bot}$,  characterizing the twofold periodicity of the collinear AF order.
These sets present symmetries, reflecting in a zero total magnetization at $T=0$.
When thermal fluctuations are taken into account, an unexpected (small) finite magnetization can be observed before reaching $T_{N}$ (see the top inset in Fig. \ref{fig:mag}). This can be an effect of the cluster chosen, in which the corner sites of cluster are more affected by mean-fields than the others, leading to a slight asymmetry in the sublattice local magnetizations. This effect is expected to disappear if larger clusters are considered.
The isotropic system can also be oriented in the $z$ direction, exhibiting the same local magnetization behavior as discussed. However, when a weak magnetic field is applied, the orientation of the system becomes transverse to the field direction, which is the choice presented in Fig. \ref{fig:mag}.

In addition, we evaluate  the magnetic susceptibility $
    \chi(T) =d\braket{M^{z}}/dh^{z} $
to compare it with the experimental measurement reported in Ref. \cite{Yamaguchi}. 
Our theoretical result for $1/\chi(T)$ at high temperatures is exhibited in the inset of Fig. \ref{fig:mag}. 
The inverse magnetic susceptibility exhibits a linear behavior at high temperatures, following the Curie-Weiss law with Weiss temperature of $\theta_{W}^{CMF}/J_{1}=-0.28$ estimated using a linear fit (red line). 
The obtained $\theta_{CW}^{CMF}=-0.28J_{1}= -6.41 K$ (assuming $J_1=22.9K$) represents a good estimate, within the CMF approach, for the experimental value $\theta_{CW}^{exp}=-8.2K$ \cite{Yamaguchi}.
A potential improvement to our quantitative results, aiming to approach the experimental value in Ref. \cite{Yamaguchi}, could involve increasing the cluster size to capture more exact correlations within the CMF scheme. 
However, this leads to numerical exact diagonalization difficulties associated with the exponential growth of the Hilbert space dimension.
Furthermore, considering that the present 8-site CMF method already captures qualitative behavior, a larger cluster may only impact a negligible quantitative improvement.

\section{SUMMARY AND CONCLUSION}\label{sec:summary}

We employed the CMF method to investigate the isotropic and anisotropic Heisenberg model with competing interactions on the square lattice under an external magnetic field  for the compound $\mathrm{(o-MePy-V)PF_{6}}$, where six different FM and AF exchange couplings are considered between the nearest neighbors.
Our results elucidate the ground-state and finite-temperature behavior of the system, revealing how the interplay of frustration, quantum fluctuations, and anisotropy shapes its magnetic and thermodynamic properties.

In the ground state, the CMF method captures an unusual increase in magnetization as a function of the magnetic field. For example, magnetization exhibits a 1/2-plateau-like state in the isotropic case, as expected for the compound ($o$-MePy-V)PF$_6$. 
A collinear AF order with spins transverse to the field direction takes place, where two quantum spin chains with competing interactions are identified. The field introduces quantum fluctuations in one of these chains (chain 2) due to frustration, decreasing its average magnetic moments and driving it to a plateau-like behavior.

The  spin anisotropy introduces relevant changes to the ground state.
For example, when finite anisotropy is considered, the previously reported collinear AF phase occurs in the longitudinal direction in weakly applied fields. 
We also observe intriguing effects emerging from the interplay between frustration and field-induced quantum fluctuations, such as the formation of singlet pairs in a nonmagnetic state. 
Notably, these effects are most prominent within an intermediate anisotropy regime, where the system can stabilize an exotic phase of coexisting FE and dimerized states. Quantum fluctuations are suppressed only as the anisotropy approaches the Ising limit.

By allowing thermal fluctuations, our results also capture some interesting behaviors observed in experiments. 
The temperature dependence of the specific heat shows a sharp peak that shifts to lower temperatures as the external magnetic field increases. 
Interestingly, the local magnetizations and longitudinal susceptibility support a phase transition from a disordered phase to a collinear AF state.

The CMF approach accurately captures the ground state of the present frustrated quantum magnet, which is in very good agreement with the tensor network method.
Notably, CMF has also proven to be a useful framework for studying the physics of this frustrated system at finite temperatures, providing qualitative insights that may inspire the search for other compounds forming square lattices with anisotropic interactions. The CMF theory calls attention to its simplicity, low computational cost, and flexibility in exploring large parameter spaces and complex lattice geometries, where tensor network methods could be computationally demanding due to limitations in bond dimension \cite{ueda}.
In addition, this work unveils the possibility of unconventional magnetism in certain verdazyl-based charge-transfer salts, highlighting the importance of the interplay between frustration and quantum fluctuations in spin systems.

\section{ACKNOWLEDGMENT}\label{sec:acknowledgment}
This work was supported by Brazilian agencies Conselho Nacional de Desenvolvimento Científico e Tecnológico (CNPq), process 165330/2023-6 and Coordenação de Aperfeiçoamento de Pessoal de Nível Superior (Capes). FMZ and LMR also acknowledge support from the Fundação de Apoio ao Desenvolvimento do Ensino, Ciência e Tecnologia do Estado de Mato Grosso do Sul (Fundect).

\bibliography{bib.bib}

\end{document}